\documentstyle[aps,prl,multicol,epsf]{revtex}
\input epsf
\begin{document}
\draft

\title{Liquid Conservation and Non-local Interface Dynamics in
Imbibition \\ 
}
\author{M.\ Dub\a'e$^{1,3,*}$, M.\ Rost$^{1,3,*}$, K.\ R.\ Elder$^2$, 
M.\ Alava$^3$, S.\ Majaniemi$^{1,3}$, and T.\ Ala-Nissila$^{1,3,4}$}
\address{
$^1$Helsinki Institute of Physics, P.O.\ Box 9, FIN--00014
University of Helsinki, Finland \\
$^2$ Department of Physics, Oakland University, 
Rochester, MI 48309--4487, USA \\
$^3$ Helsinki University of Technology,
Laboratory of Physics, P.O.\ Box 1100, FIN--02015 HUT, Finland \\
$^4$Department of Physics, Brown University, Providence
RI 02912--1843, USA
}

\date{\today}

\maketitle

\begin{abstract}

The propagation and roughening of a liquid-gas interface
moving through a disordered medium under the influence of capillary forces 
is considered.  
The system is described by a phase-field model with conserved 
dynamics and spatial disorder is introduced through a quenched random field.  
Liquid conservation leads to slowing down of the average interface position 
$H$ and imposes an intrinsic correlation length 
$\xi_\times \! \sim \! H^{1/2}$ on the spatial fluctuations of the interface.
The interface is statistically self affine in 
space, with global roughness exponent 
$\chi \! \simeq \! 1.25$ and exhibits anomalous scaling.

\end{abstract}

  
\begin{multicols}{2}

The dynamics of an invading liquid front in a disordered medium involves
many problems of interest in statistical physics such as transport and
diffusion in random media \cite{Bouchaud_93,Scheidegger_57},
roughening of a moving interface \cite{Krug_97}, and pinning
\cite{Kardar_98} if the interface stops.  Many experiments have 
been conducted on such phenomena\cite{Barabasi_95}, 
examples include Hele-Shaw 
cells\cite{Hele-Shaw} and the {\it spontaneous imbibition} of water in paper
\cite{Buldyrev_92,Family_92,Amaral_94,Horvath_95,Zik_97}.
In the latter case, capillary forces drive the liquid until balanced 
by loss of water by evaporation or hydrostatic pressure. Often 
local theories such as directed percolation
depinning \cite{Buldyrev_92,Amaral_94} have 
been successful in describing the properties of such pinned 
fronts \cite{Kardar_98}.

	Local interface theories cannot however provide a complete 
description of imbibition since the transport of liquid to the front 
from the reservoir must be taken into account.   In particular,
this leads to a slowing down of the invading front
even {\em without} evaporation or gravity 
\cite{Washburn_21}.   
In this letter a model of imbibition that explicitly addresses this issue 
through the inclusion of liquid conservation, is introduced. 
This is achieved by a generalised
Cahn-Hilliard equation with a non-equilibrium boundary condition that 
couples the system to a liquid reservoir.
Two different cases are considered: a liquid front invading a 
disordered medium without gravity or evaporation, and a steady 
state front maintained  at a constant distance from the 
liquid reservoir as done in the experiments of 
Horv\'ath and Stanley \cite{Horvath_95}.

	An obvious consequence of liquid conservation is that 
the interface dynamics is nonlocal.  One manifestation of 
this non-locality is that the average interface height $H$ grows 
as $t^{1/2}$ as might be expected by Washburn's 
equation \cite{Washburn_21} (the velocity, $\dot{H} \sim 1/H$).
The interface is found to be superrough, with global roughness 
exponent $\chi \! \simeq \! 1.25$, and exhibits anomalous scaling
\cite{Krug_97,Lopez_97}.  
Interface fluctuations saturate at a 
correlation length $\xi_\times$,
which is controlled by the  average 
interface height such that $\xi_\times  \! \sim \! H^{1/2}$. 
This implies that the average interface width $W \! \sim
\! \xi_\times^\chi$ \cite{note_overhangs}.  The relationship 
between $\xi_\times$ and $H$ is a direct consequence of 
the conservation law and arises from the interplay between
driving force and curvature.   No such relation occurs in local 
models \cite{Barabasi_95}.  The temporal correlations of the 
interface qualitatively agree with the experiments of Ref.\ 
\cite{Horvath_95}.  It is further found that the front
moves by avalanches, as reflected in the behaviour of the higher
moments of the temporal correlations
\cite{LeschhornTang}. 

 {\em Phase field model of imbibition:}  To describe the liquid/gas 
system a locally conserved field  $\phi ({\bf x},t)$ is introduced 
such that $\phi=+1$ and $-1$ are the liquid and gas phases 
respectively.  The free energy functional of the system can 
then be written, ${\cal F} \{ \phi\} = \int d{\bf x} 
[ (\nabla \phi)^2 \! /2 + V(\phi)]$, where $V({\bf x},\phi) \equiv 
- \phi({\bf x},t)^2/2 +  \phi({\bf x},t)^4/4 - \alpha({\bf x}) 
\phi({\bf x},t)$.   The liquid/gas interface is thus sustained 
in the standard manner by the gradient energy term 
and double well structure of $V$.  The variable $\alpha({\bf x})$ 
accounts for the random nature of the medium and has 
correlations; $\langle \alpha({\bf x}) \rangle \! = \! \bar
\alpha$, and $\langle \alpha({\bf x}) 
\alpha({\bf x'})\rangle \! - \! \bar \alpha^2 \! = \! (\Delta \alpha)^2
\delta({\bf x} \! - \! {\bf x'})$.  The propensity of the medium 
to absorb liquid can be controlled by $\alpha$ since it determines the 
local chemical potential ($\mu = -\delta F/\delta \phi$).
The dynamical evolution of $\phi$ is determined by a continuity equation 
$\partial_t \phi \! + \! \nabla \! \cdot \! {\bf j} = 0$, 
where ${\bf j} ({\bf x},t) = - \nabla \mu ({\bf x},t)$. 
The resulting equation of motion, i.e., 
\begin{equation}
\label{pf_eq}
\partial_t \phi = - \nabla^2 \left[ \nabla^2 \phi + \phi - \phi^3 +
\alpha({\bf x}) \right]
\end{equation}
is a slightly modified Cahn--Hilliard equation \cite{Cahn_58}. 
The absorption of liquid by a dry random medium can be modeled 
with the appropriate boundary conditions.  To reflect the 
presence of an infinite resevior at $y \leq 0$ the chemical 
potential at $y=0$ is fixed to be zero.  To induce the propagation of the 
front, the top end of the system is kept ``dry'' 
(i.e., $\phi(y \! \to \! \infty) = -1$) and $\bar{\alpha} > 0$.  
Equation (\ref{pf_eq}) can be augmented to include evaporation by 
adding a term proportional to $(\phi \! + \! 1)$.

This model was specifically constructed to incorporate several 
essential physical features of imbibition.  More precisely 
the model incorporates the conserved nature of the fluid which 
leads directly to a Gibbs-Thomson boundary condition 
on the liquid front and an average front velocity that 
decreases in time \cite{note1}.  To see these properties and gain more insight 
into the model it is useful to project out the 
equation of motion of the interface using standard projection operator 
methods \cite{Kawasaki_82,Dube_Long}.   Expanding to lowest order in the
interfacial curvature (${\cal K}$) and $\bar{\alpha}$ \cite{lowest} gives;
\begin{equation}
\label{non-local}
\Delta \phi \int_{-\infty}^{\infty} dx' {\cal G} (x,h|x',h')
\dot{h}'= \eta (x,h)  + \sigma {\cal K}
\end{equation}
where ${\cal G}$ is the half-plane Green's function given by 
${\cal G}(x,y|x',y') = \frac{1}{4\pi} \ln \, \left( \, \frac{(x-x')^2 +
(y-y')^2}{(x-x')^2+(y+y')^2} \, \right)$ which reflects a broken 
translational symmetry along the $y$-axis due to the reservoir.
The interface is defined by 
the single-valued function $h(x,t)$, 
$\sigma \equiv \int dy ( \phi_0')^2$ is the surface tension, 
$\phi_0$ is the one-dimensional kink solution of
Eq.\ (\ref{pf_eq}) and $\Delta \phi = 2$ is the miscibility gap. 
The quenched noise can be written
$\eta(x,h) \equiv \int dy \phi_{0}'(y-h(x,t)) \, \alpha(x,y)
\sim 2 \alpha (x,h)$ in the sharp interface limit. Similar non-local
equations also arise in the context of Laplacian fluid flow
\cite{Krug_91} and step growth \cite{Bales_90}. Previous theoretical
work on
equations for interfaces at the depinning transition \cite{Kardar_98}
and nonlinear equations with long range kernels
\cite{Mukherji_97} does not
apply to the situation encountered here.

The Gibbs-Thomson condition, $\mu|_{int}  \sim {\cal K}$ can be
immediately obtained from Eq.\ (\ref{non-local}) in the limit
$\dot{h}=0$ since $\eta$ is the chemical potential at the interface.
In addition since  $\phi$ is a conserved field it is easy to show that
the normal interface velocity is proportional to the gradient of $\mu$ 
at the interface, i.e., $\Delta \phi v_n = - \partial_n \mu |_{int}$.
Further information can be obtained by linearising in $h$ to obtain;
\begin{equation}
\label{effint}
\dot{h_k} \left( 1 \right.- e^{-2 \vert k \vert H} \left.\right)
+  |k| \dot H \; h_k \left( 1 + e^{-2 \vert k \vert H}
\right) = 
|k| \left(\eta_k -  \sigma k^2 h_k \right), 
\end{equation}
where $h_k$ are the Fourier components of $h$ and $H=h_0$ is the 
average interface position \cite{local_nonlocal}.
Equation (\ref{effint}) can now 
be used to obtain an expression for the average interface position 
(Washburn's equation) and lateral correlation length as 
follows.

 {\em Washburn's equation:} In the limit $k \rightarrow 0$, Eq.\
(\ref{effint}) reduces to $\dot{H} \! = \! \bar \alpha/(2H)$.  Thus a
planar interface, without evaporation is never pinned and moves as
$H(t) = (\bar \alpha t)^{1/2}$.  This is a standard result known as
Washburn's equation \cite{Washburn_21} and can be understood as
follows.  In the absence of a liquid reservoir $\mu  \! = \! - \bar
\alpha$ throughout the system.  When a reservoir is included at the
lower boundary $\mu$ is set to zero at $y=0$.  For a slowly moving front
$\mu(y)$ satisfies a diffusion equation in the bulk phases with 
quasi-stationary solution $\mu \approx -\bar{\alpha} y/H$ for $y \le \! H$, 
and $\mu (y) = - \bar \alpha$ for $y \ge H$.  This result is equivalent
to Washburn's equation since $\dot{H} = v_n  = - 
(\partial_n \mu |_{int})/\Delta \phi$.  In essence the gradient in
$\mu$ creates a current from the reservoir towards the interface,
causing it to advance. 

 {\em Lateral correlation length:} Another useful result coming from
Eq.\ (\ref{effint}) is the existence of a lateral length scale
separating two different modes of damping of the interface
fluctuations. Long wavelength fluctuations are damped by the advancing
motion of the front (i.e., $|k| \dot{H} h_k$), while short wavelength
fluctuations are damped by surface tension (i.e., $\sigma |k|^3
h_k$). The length scale separating these two modes $\xi_{\times} \!
\sim \! (\sigma/\dot{H})^{1/2} \! = \! (\sigma H/\bar \alpha)^{1/2} $
is closely related to the
Mullins-Sekerka instability of driven Laplacian fronts
\cite{Mullins_64}, although the situation is reversed here.
Because fluid is transported towards the front from behind, advanced
(retarded) parts receive less (more) mass than the average and the
front is {\it stabilised} at long length scales. 

	This result can be understood in more physical terms as follows.
Due to the Gibbs--Thomson effect a local ``bulge'' of vertical extent $W$ 
and lateral size $\xi$ alters the chemical potential by $\Delta \mu 
\simeq \sigma W/\xi^2$. On the other hand, the average gradient
in $\mu$ in the bulk liquid induces a difference $\Delta \mu \simeq
\bar \alpha W /H$ across a vertical distance $W$. These two differences
balance each other at a length given by
\begin{equation}
\label{xi_cross}
\xi_\times \; \simeq \; \sqrt{\sigma H/\bar \alpha}.
\end{equation}
By this mechanism, one expects the front to be smoother on length scales
larger than $\xi_{\times}$ as compared to smaller scales.  While these 
interface results provide useful insight into the front propagation the
behavior was examined in more detail by direct numerical simulation.

\begin{figure}
\narrowtext
\epsfxsize=3.2in \epsfysize=3.2in
\epsfbox{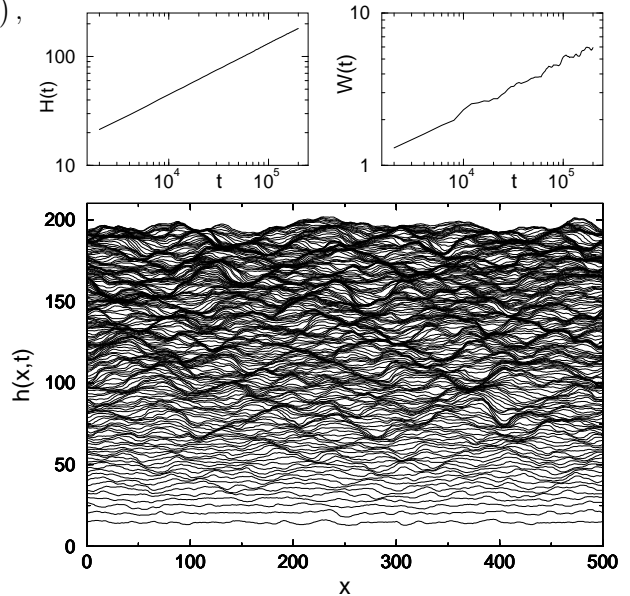}
\caption{Configurations of the rising interface at time intervals
$\Delta t=10^3$. For this system, the average height $H \sim t^{0.49}$, 
and the width $W(t) \sim t^{0.32}$, according to least-square fits.
}
\label{rise}
\end{figure}
 {\em Numerical integrations:} Equation (\ref{pf_eq}) was integrated
on a square grid with spacing $\Delta x \! = \! 1$ using various lateral
system sizes $L_x$.  The disorder $\alpha({\bf x})$ was a random variable 
on each grid point. Different distributions (exponential, Gaussian, 
{\it etc.}) gave the same results. The interface position $h(x,t)$ 
was determined by a linear interpolation of the zero of the phase field,
$\phi(x,h(x,t)) \! = \! 0$, with overhangs (hardly present here)
cut off. 

Fig.\ \ref{rise} shows the rise of an initially flat 
front. Fronts are shown at equal time intervals, displaying 
Washburn behavior $H^2 (t) \! = \! \bar{\alpha} t$, also shown 
in the Figure. The vertical and lateral extent of interface 
fluctuations increase with height. The early time behavior of 
the width $W(t) = \langle \overline{(h(x,t) -
\bar{h})^2} \rangle^{1/2}$, where overbar denotes a
system average and the brackets average over the disorder, is also
shown in Fig.\ \ref{rise}. It is found that $W(t) \sim 
t^{\beta}$ with $\beta = 0.32 \pm 0.02$. 

In a recent experiment of paper wetting \cite{Horvath_95},
a steady state situation was reached by pulling the paper
sheet at a constant velocity ${\bf v} \! = \! -v {\bf \hat y}$ towards
the reservoir. To model this case, Eq.\ (\ref{pf_eq}) can be written
as
\begin{equation}
\label{horvath_eq}
\partial_t \phi = - \nabla^2 \left[ \nabla^2 \phi + \phi - \phi^3 +
\alpha({\bf x}-{\bf v}t) \right] + {\bf v} \cdot 
{\mbox {\boldmath $\nabla$}} \phi,
\end{equation}
keeping the original boundary conditions. In this case, the interface
remains at a {\em fixed} average height $H = \bar{\alpha}/2 v$, at
which a freely rising interface would have velocity $v$.

\begin{figure}
\narrowtext
\epsfxsize=8.0cm \epsfysize=6.0cm
\epsfbox{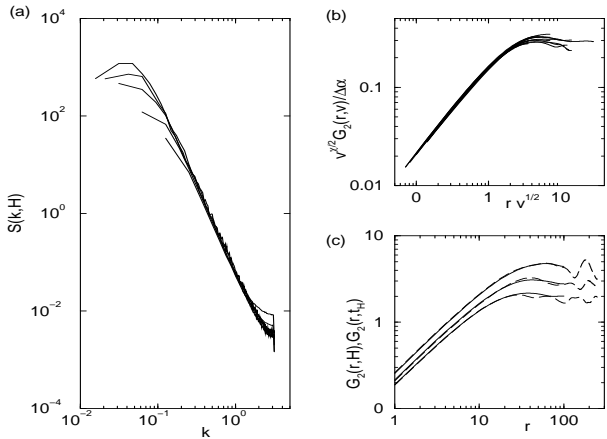}
\caption{(a) Structure factors for systems of height $H \! = \!
50,100,150,200$ and $L_x \! = \! 2 H$. (b) Correlation functions $G_2
(r,H)$, scaled according to Eq.\ \protect (\ref{g_scaling}) for
various $v \! \sim \! H^{-1/2}$ and $\Delta \alpha$. (c) Comparison of
the correlation functions in the steady state, $G_2 (r,H)$ (solid
lines) for $H \! = \! 25,50,100$ and of the rising case $G_2 (r,t_H)$
(dashed lines), at times $t_H \! = \! H^2/\bar \alpha$.} 
\label{spatial}
\end{figure}
Equation (\ref{horvath_eq}) was also integrated numerically. The role of
$\xi_\times$ as a maximal range of correlated roughness
becomes apparent in the steady-state structure factor $S(k,H) =
\langle \overline{\vert h(k,t) \vert^2} \rangle$ and 
spatial height difference correlation function 
$G_2(r,H) = \langle \overline{(h(x+r,t) - h(x,t))^2} \rangle^{1/2}$
\cite{argument_H}, both shown on Fig.\ \ref{spatial}. 
The lateral fluctuations of the interface are correlated only 
up to $\xi_\times \! < \! L$. The global roughness exponent 
$\chi$ is taken from the power law decay
of $S(k \! > \! \xi_{\times}^{-1},H) \! \sim \! k^{-(2 \chi + 1)} =
k^{-3.5 \pm 0.2}$, yielding $\chi \simeq 1.25$. The
relation $\xi_\times \! \sim \! v^{-1/2} \! \sim \!(H/\bar \alpha)^{1/2}$ is 
confirmed in Fig.\ \ref{spatial}, were the spatial correlation 
function is rescaled as
\begin{equation}
\label{g_scaling}
G_2(r,H) = \Delta \alpha \; v^{-\chi/2} \; g \left(
r \, v^{1/2} \right),
\end{equation}
for different $v$ (i.e., different $H$) and $\Delta \alpha$. 
The scaling function
$g(u)$ is constant for $u \gg 1$ and $g(u) \sim u^{\chi_{loc}}$ 
if $u \ll 1$, with $\chi_{loc} \simeq 1$. 
The interface is thus superrough, and $G_2(r,H)$ shows anomalous scaling 
in the sense that $\chi_{loc} \! \! < \! \chi$. $G_2(r,H)$ 
increases with $H$ for all $r$ \cite{Krug_97,Lopez_97}.

{\em Quasi-stationary rise:} The numerical data indicate that 
spatial correlations in the interface position are the {\em same} 
for both sets of simulations.  This can be seen in Fig.\ \ref{spatial}
where steady state correlation functions at fixed heights $H$ 
are compared with the correlation functions of a 
freely rising front, obtained from Eq.\ (\ref{pf_eq}), 
at times $t_H = H^2/\bar{\alpha}$. This implies 
that the interface is always in a quasi-stationary state in the 
freely rising case.  The scaling form, Eq.\ (\ref{g_scaling}),  
is thus also valid for the freely rising front provided that the 
time dependence $H(t) \! = \!
(\bar \alpha t)^{1/2}$ is used, such that $\xi_\times \! \sim \!
(t/\bar{\alpha})^{1/4}$ \cite{Dube_Long}. This indicates that the 
dynamical correlation length $\xi_t \! \sim \! t^{1/z}$ increases at
a rate faster or equal to $t^{1/4}$, so that the
interface fluctuations can always catch up with the maximal
correlation range. This also implies $W \! \sim \!
\xi_\times^\chi \! \sim \! t^{\chi/4} = t^{\beta}$, 
with $\beta \simeq 0.31$, in agreement with $W(t)$ 
shown in Fig.\ \ref{rise}.

\begin{figure}
\narrowtext
\epsfxsize=8.0cm \epsfysize=6.0cm
\epsfbox{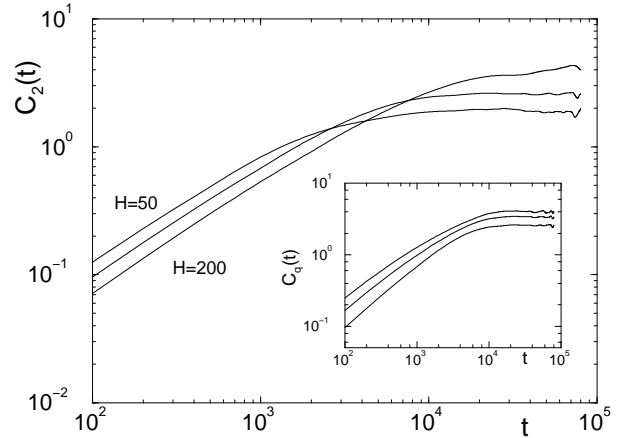}
\caption{Steady state temporal correlation functions for
systems of size $L_x = 400$, at heights $H=50,100,200$.
In the inset, the moments $q=2,4,6$ (from bottom to top) for $H=100$
are compared. $C_2$ has exponent $\beta_2 = 0.85$, while
for the higher moments, the effective exponents
 $\beta_4 \approx 0.76$ and $\beta_6 \approx 0.69$.
}
\label{temporal}
\end{figure}
{\em Temporal correlations:} To compare with the experimental work
of Ref.\ \cite{Horvath_95}, the temporal correlations in the steady
state, obtained from Eq.\ (\ref{horvath_eq}), were analysed 
through the function $C_q(t,H) = \langle \overline{\vert
h(x,t+s) \! - \! h(x,s) \vert^q} \rangle^{1/q}$ also taking higher
moments $q \! > \! 2$ into account. The early time logarithmic slopes
of the higher moments ($\beta_q$) {\em decrease} with $q$,
as shown in the inset of 
Fig.\ \ref{temporal}, 
in contrast to the spatial $G_q$ for which $\chi_q$ is 
independent of $q$.
This lack of scaling reflects the propagation
of the front by avalanches (easily visible in Fig.\ \ref{rise}),
also found for local interfaces at the depinning transition 
\cite{LeschhornTang,Dube_Long}.

Nevertheless, our model agrees with the experiments in three ways,
visible in the main part of Fig.\ \ref{temporal}: {\it (i)} the
$\beta_q$ are independent of $H$; {\it (ii)} the late time saturation
level of $C_q$ increases with $H$ indicating larger overall roughness,
and  {\it (iii)} $C_q(t,H)$ at fixed small $t$ decreases with $H$,
indicating faster intrinsic avalanche velocity for smaller
$H$. Quantitatively our result $\beta_2 \! = \! 0.85\pm 0.04$ differs
from the experimental value $0.56$, but so does the average
interface velocity $v \! \sim \! H^{-1.6}$ in Ref. \cite{Horvath_95}
compared to Washburn behavior in our model \cite{betafoot}. 
Since water flow dynamics
in real paper can be complicated by various phenomena such as e.g.\
fiber swelling \cite{Dube_Long}, the disagreement is not too surprising
and should be clarified in future investigations.

{\em Conclusion:} We have constructed a simple but
flexible model for spontaneous imbibition of a liquid from a reservoir
into a disordered medium. Conservation of liquid leads to Washburn's relation 
for the average height $H \! \sim \! t^{1/2}$ \cite{Washburn_21}, a 
behavior which does not occur naturally in local interface models.
The interface dynamics becomes {\em nonlocal}, and an intrinsic
length scale $\xi_{\times}$ emerges as given by Eq.\ (\ref{xi_cross}).
This length scale is a distinct prediction of the model, and intimately
relates the fluctuations of the interface with its average progression.
The existence and dynamical increase of this region should be 
experimentally easier to observe than precise values of scaling
exponents on scales $r < \xi_{\times}$. From the model, spatial
scaling is found to be anomalous with a global roughness exponent
$\chi \! \simeq \! 1.25$, similar to what has been seen in other
models of driven interfaces in disordered media
\cite{Leschhorn_93}. The growth exponent $\beta$ is explained in terms
of $\xi_{\times}$ and $\chi$. 
Acknowledgements: This work has in part been supported by the Academy
of Finland and a Research Corporation Grant, No. CC4787 (KRE).

\end{multicols}


\begin{thebibliography}{99}
\bibitem{note_overhangs} We only consider the case of small amplitudes
in interface fluctuations. In the presence of substantial overhangs
one expects $W \! \sim \! \xi_\times$.
\bibitem{lowest} The expansion is around a stationary planar interface.
In such an expansion the small parameters are the dimensionless 
quantities, $\zeta{\cal K}$ and $\zeta \bar{\alpha}/H$, where 
$\zeta$ is the interfacial thickness.
\bibitem{local_nonlocal} Equation (\protect \ref{effint}) acquires a
local form in the limit $\vert k \vert H \ll 1$. Two points on
the interface separated by a distance larger than $H$ are decoupled.
\bibitem{betafoot} $\beta_2$ is {\em not} the
exponent $\beta$ of the early time increase of the RMS width
in the freely rising case, contrary to the suggestion made 
in Ref. \cite{Horvath_95}, since the latter is determined 
through $\xi_{\times}$.
\end{thebibliography}
\end{document}